\documentclass{article}
\usepackage{spconf,amsmath,graphicx,enumitem,IEEEtrantools,pifont,booktabs,verbatim}

\title{Time-domain Speech Enhancement Assisted by \\ Multi-resolution Frequency Encoder and Decoder}
\name{Hao Shi$^{1}$,~Masato Mimura$^{1}$,~Longbiao Wang$^{2}$,~Jianwu Dang$^{2}$,~Tatsuya Kawahara$^{1}$}
\address{
$^{1}$Graduate School of Informatics, Kyoto University, Kyoto, Japan\\
$^{2}$Tianjin Key Laboratory of Cognitive Computing and Application, \\
College of Intelligence and Computing, Tianjin University, Tianjin, China\\
}
\begin{document}
%
\maketitle

\begin{abstract}
\small 
Time-domain speech enhancement (SE) has recently been intensively investigated. 
Among recent works, DEMUCS \cite{Dfossez2020} introduces multi-resolution STFT loss to enhance performance. 
However, some resolutions used for STFT contain non-stationary signals, and it is challenging to learn multi-resolution frequency losses simultaneously with only one output. 
For better use of multi-resolution frequency information, we supplement multiple spectrograms in different frame lengths into the time-domain encoders. 
They extract stationary frequency information in both narrowband and wideband. 
We also adopt multiple decoder outputs, each of which computes its corresponding resolution frequency loss. 
Experimental results show that (1) it is more effective to fuse stationary frequency features than non-stationary features in the encoder, and (2) the multiple outputs consistent with the frequency loss improve performance. 
Experiments on the Voice-Bank dataset show that the proposed method obtained a 0.14 PESQ improvement. 
\end{abstract}
\section{Introduction}
\label{sec:intro}
{\small
Speech enhancement (SE) has been extensively studied because noise often corrupts speech signals collected in real-world scenarios \cite{6732927, chen2023metric, 10.1007/978-3-319-22482-4_11}, 
which significantly degrades the performance of speech applications \cite{chen2022self, 45168, 6639038, 6843279, 9747755, chen2022leveraging, 9689650}. 
SE aims to recover speech components from noisy signals \cite{9054661}.  
Deep learning-based SE \cite{9103053, 6932438} methods have been shown to perform better than traditional methods \cite{5745591,397090}. 
Supervised learning-based SE can be classified into frequency-domain \cite{9103053}, and time-domain \cite{8683634} methods. 
Frequency-domain SE that uses only magnitude information has been mainly studied because it presumes that the human ear is less sensitive to phase information than magnitude information. 
This issue was supplemented and corrected by subsequent studies \cite{1451871}. 

Recently, SE systems that process magnitude and phase information simultaneously achieves impressive performance \cite{Yin_Luo_Xiong_Zeng_2020, 8682834}. 
There are two approaches to handle phase information: enhancement in complex-domain \cite{Yin_Luo_Xiong_Zeng_2020, 8682834} and time-domain \cite{Dfossez2020}. 
Complex-domain SE \cite{8682834, Hu2020} processes the Fourier transform's real and imaginary parts. 
Time-domain SE \cite{Dfossez2020, 9689572} directly inputs the time-domain waveform and outputs enhanced features. 

Among time-domain SE models, DEMUCS \cite{Dfossez2020} has demonstrated state-of-the-art performance. 
It is based on the standard U-Net \cite{10.1007/978-3-319-24574-4_28} structure and optimized by minimizing the L1 regression loss and supplemented by multi-resolution spectrogram domain losses \cite{9053795}. 
DEMUCS exploits frequency-domain information through the spectrogram-domain loss, which significantly improves the stability of the model training. 
Furthermore, different from other time-domain models \cite{8701652, PASCUAL201910, 8683634}, DEMUCS introduces upsampling \cite{1172555}, and downsampling \cite{1172555} processing based on sinc interpolation before the encoder and after the decoder, respectively, and the interpolation of the redundant information can alleviate information loss or distortion \cite{https://doi.org/10.48550/arxiv.2201.06685} caused by SE.

DEMUCS still has two drawbacks: 
(1) Some of the supplemented frequency-domain information contains non-stationary signals. 
Speech signals can be regarded as short-term stationery with an interval between 10ms and 30ms \cite{862099}. 
The Fourier transform presupposes that the signal is stationary \cite{1164317}. 
However, in addition to 32ms, DEMUCS also adopts STFT of 64ms and 128ms. 
(2) It needs to learn the frequency loss of different resolutions simultaneously with one output. 
Multiple learning targets with a single output make training the neural network difficult.

\begin{figure*}[h]
	\centering 
	\includegraphics[width=1.\textwidth]{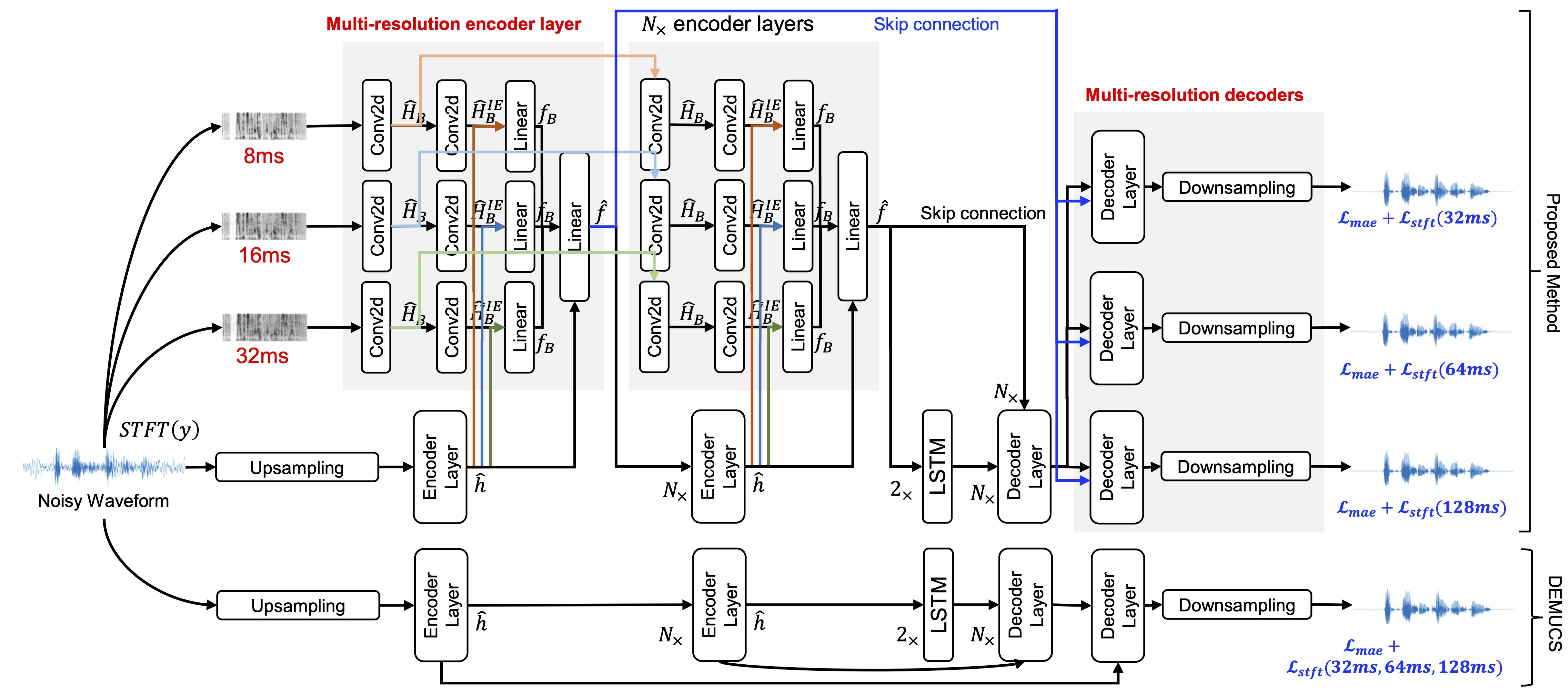}
	\vspace{-20pt}
	\caption{\small Flowchart of the proposed method and DEMUCS. The proposed multi-resolution encoder and decoders are highlighted in the flowchart.} 
	\vspace{-10pt}
	\label{fig1}
\end{figure*}

In this study, we investigate the better use of multi-resolution frequency information from the perspective of the encoder and decoder, respectively. 
First, 
    instead of using non-stationary frequency information in the output, we incorporate multi-resolution stationary frequency-domain information into the time-domain SE encoder layer by layer. 
    The multi-resolution spectrograms are supplemented to provide frequency domain information. 
    According to the length of framing time, spectrograms can be divided into wideband and narrowband \cite{150375}. 
    These two kinds of spectrograms are much different and show a certain complementarity: 
    Wideband (about 3ms length of framing time) spectrogram can capture the rapid amplitude changes \cite{5213512} and clear speech formant information; 
    Narrowband (about 20ms length of framing time) spectrogram has better spectral resolutions and captures harmonics information. 
    Furthermore, the SE system \cite{9053795} trained with multi-resolution information outputs significantly higher perceived quality in mean opinion score (MOS). 
    Considering these, we incorporate frequency information of 8ms, 16ms, and 32ms into the model. 
Second, 
    we propose using multiple time-domain decoders by downsampling, each corresponding to only one resolution of frequency-domain loss.

In the following sections, we will introduce related work in Section 2. 
We will introduce the proposed method in Section 3. 
In Section 4, the experimental settings and results will be introduced. 
The conclusion will be introduced in Section 5. 
}

\section{DEMUCS}
{

\small
The time domain SE directly inputs a  noisy speech waveform $y$ and outputs enhanced waveform $\hat{x}$: 
\begin{equation}
	\hat{x} = \mathcal{N}(y)
	\label{eq5}
\end{equation}
The mean absolute error between $\hat{x}$ and the original signal $x$ is the common loss function for training time-domain SE: 
\begin{equation}
	\mathcal{L}_{mae} = \frac{1}{T}||\hat{x} - x||_F^1,
	\label{eq6}
\end{equation}
where $T$ is the time points in the waveform.

\textbf{DEMUCS} is one of the time-domain SE shown in Fig.\ref{fig1}. 
It is based on the standard U-Net structure. 
It contains five encoder layers, two Long Short-term Memory (LSTM) layers, and five decoder layers. 
During training, in addition to the time-domain loss in Eq.~(\ref{eq6}), DEMUCS also introduces the following two frequency-domain losses: 
\begin{equation}
	\begin{aligned}
		& \mathcal{L}_{stft} = \mathcal{L}_{sc} + \mathcal{L}_{mag} \\
		& \mathcal{L}_{sc} = \frac{ |||STFT(\hat{x})| - |STFT(x)|||_F^1 }{|STFT(x)|} \\
		& \mathcal{L}_{mag} = \frac{1}{T}||\text{log}|STFT(\hat{x})| - \text{log}|STFT(x)|||_F^1 \\
	\end{aligned}
	\label{eq7}
\end{equation}
The final multi-resolution loss function of DEMUCS is:
\begin{equation}
	\mathcal{L}_{demucs} = \alpha \mathcal{L}_{mae} + (1 - \alpha) \sum\limits_{r=1}^{R} (\mathcal{L}_{stft}(r))
	\label{eq8}
\end{equation}
where $R$ represents the multi-resolution number. 
In the conventional standard DEMUCS, $R = 3$ and its STFT points are \{32ms, 64ms, 128ms\}, the hop size are \{3.125ms, 7.5ms, 15ms\}, and the window (Hanning window) length are \{15ms, 37.5ms, 75ms\}. 
}

\section{Proposed Method}
{
\small
Although the STFT loss introduced in DEMUCS shows a significant improvement, it still has two problems: framing lengths \{64ms, 128ms\} are non-stationary for speech signals, and single output for multiple resolution STFT information may increase the learning burden of the neural network due to the mismatch. 
Fig.~\ref{fig1} shows a flowchart of the proposed method. 
We address the above two issues from the following two aspects.

\subsection{Fusing Frequency Information in Encoder}
Instead of introducing STFT information in the loss calculation, multi-resolution stationary frequency information is incorporated into the encoder layer by layer. 
Time-domain branch is processed:
\begin{equation}
    \begin{aligned}
	& \hat{h} = GLU(Conv1d((ReLU(Conv1d(h)))))
	\end{aligned}
\label{eq_pe_input_time}
\end{equation}
where $h$ represents the time information from the output of the previous encoder layer or the original time-domain input feature. 
$\hat{h}$ is a hidden representation obtained by convolutional processing in Eq.~(\ref{eq_pe_input_time}).

Three different window-size spectrograms are adopted to provide stationary multi-resolution frequency information. 
Spectrograms can be divided into wideband and narrowband spectrograms according to the number of STFT points with certain information complementarity. 
Taking into account the short-term stability of the speech signal, we choose \{8ms, 16ms, and 32ms\} with \{4ms, 8ms, 16ms\} hop size and \{8ms, 16ms, 32ms\} window length as frequency input features.

Frequency information in each encoder layer is processed as follows: 
\begin{equation}
    \begin{aligned}
	& \hat{H_B} = ELU(BatchNorm2d(Conv2d(H_B))), \\
    \end{aligned}
\label{eq_pe_input}
\end{equation}
$H_B$ represents the $B-$th frequency information from the output of the previous encoder layer or the original frequency feature, $B$ is among \{32ms, 16ms or 8ms\}. 
$\hat{H_B}$ is a hidden representation obtained by convolutional processing in Eq.~(\ref{eq_pe_input}). 
$\hat{H_B}$ is adopted as the $B$-th frequency input to the next encoder layer. 
Furthermore, $\hat{H_B}$ is used as auxiliary information to improve the time-domain SE branch.

In order to extract a more suitable feature representation from frequency information, the $\hat{H_B}$ is processed by two more convolutional processing: 
\begin{equation}
	\begin{aligned}
		& H_B^{IE} = ELU(BatchNorm2d(Conv2d(\hat{H_B}))), \\
		& \hat{H_B^{IE}} = ELU(BatchNorm2d(Conv2d(H_B^{IE}))), \\
	\end{aligned}
	\label{eq_pe_ei}
\end{equation}
where $\hat{H_B^{IE}}$ is the extracted information from $\hat{H_B}$.

Finally, the frequency information is incorporated into the time-domain branch as follows: 
\begin{equation}
	\begin{aligned}
		& f_B = Linear(Concat(\hat{h}, \hat{H_B^{IE}})) \\
		& \hat{f} = Linear(ReLU(Linear(\hat{h} + f_{8} + f_{16} + f_{32})) \\
	\end{aligned}
	\label{eq_pe_fuse}
\end{equation}
$\hat{f}$ and $\hat{H_B}$ are the time and frequency domain outputs of the encoder layer, respectively. 
$\hat{f}$ is also adopted as skip connection information and is input into the corresponding decoder layer. 
We refer to the model with multi-resolution frequency encoder as \textbf{DEMUCS-MRE}. 
The loss function is the same with Eq.~(\ref{eq8}).

\subsection{Multiple Decoders Consistent with the Learning Targets}
In addition to the non-stationary loss issue, reducing the mismatch between the multi-resolution frequency losses and single network output is essential. 
We propose to use multiple outputs to alleviate the problem that the multiple learning targets are set for the single output. 

In the proposed method, each output only calculates one resolution STFT loss. 
In this paper, the decoder depth is five. 
So there will be three output layers in parallel after the fourth decoder layer. 
For the decoder layer, we use the same structure as DEMUCS: 
\begin{equation}
	\hat{d} = ReLU(ConvTranspose1d(GLU(Conv1d(d)))),
	\label{eq_de}
\end{equation}
where $d$ is the output of the previous decoder layer or the output of the LSTM layers. 
$\hat{d}$ is the processed output. 
The last layer of the decoder does not use the ReLU activation function.

Different decoder outputs are expected to perform better on their corresponding resolution frequency-domain information. 
Averaging multiple waveforms, especially with complementary information, can improve the enhancement performance \cite{9054661}. 
Thus, the final enhanced waveform is an average of three different outputs. 
We refer to the model that takes multiple time-domain outputs as \textbf{DEMUCS-MRD}.
The loss function is the same with Eq.~(\ref{eq8}).

}

\section{Experimental Settings and Analysis}
{
\small
All neural networks were implemented with PyTorch. 
We used the causal DEMUCS, which can be used for streaming operations. 
The detailed neural network settings can be found in this {URL}\footnote{https://github.com/hshi-speech/icassp2023/tree/main}.  

We used a public dataset synthesized from the Voice Bank corpus \cite{6709856}. 
The dataset can be accessed from this {URL}\footnote{https://datashare.ed.ac.uk/handle/10283/1942}. 
All speech data were sampled at 16 kHz. 


\subsection{Evaluation Metrics}
We used several composite measures for evaluation. 
They are obtained by linearly combining existing objective measures. 
In this paper, we used multiple linear regression analysis to form the following composite measures: $C_{sig}$ for a five-point scale of signal distortion (SIG) \cite{4389058};
$C_{bak}$ for a five-point scale of background intrusiveness (BAK) \cite{4389058};
$C_{ovl}$ for the overall quality (OVL, [1=bad, 2=poor, 3=fair, 4=good, 5=excellent]) \cite{4389058}.
The three composite measures are obtained from log-likelihood ratio (LLR) \cite{4389058}, the perceptual evaluation of speech quality (PESQ) \cite{941023}, segmental SNR (segSNR) \cite{4389058}, and weighted-slope spectral (WSS) \cite{1171512} distance. 
We also adopted the Short-Time Objective Intelligibility (STOI) \cite{5713237}. 
For all metrics, higher values indicate better performance.

\begin{table}[ht]
	\centering
	\vspace{-10pt}
	\small
	\caption{\small Comparison of different resolution loss in DEMUCS.}
	\vspace{5pt}
	\begin{tabular}{p{4.5cm}cc}
		\toprule
		System  & STOI ($\%$, $\uparrow$) & PESQ ($\uparrow$) \\
		\midrule
		DEMUCS-8ms,16ms,32ms & 94.2 & 2.79  \\
        DEMUCS-32ms       & 94.6 & 2.92  \\
        DEMUCS-32ms,64ms,128ms (Conventional) & 94.8 & 2.93 \\
		\bottomrule
	\end{tabular}
	\label{table-stationary}
	\vspace{-15pt}
\end{table}

\subsection{Effect of Different STFT losses in DEMUCS}
First, we tested different STFT losses for DEMUCS. 
We refer to standard ``DEMUCS'' as ``DEMUCS-32ms,64ms,128ms''. 
``DEMUCS-8ms,16ms,32ms'' is compared to see the effect of stationary STFT losses. 
Unexpectedly, Table~\ref{table-stationary} shows that its performance degrades compared with the standard ``DEMUCS-32ms,64ms,128ms''. 
This may be because although 64ms and 128ms are non-stationary signals for speech processing, they may benefit noise components. 
Therefore, we did not choose to change the resolutions of STFT losses in the time domain enhancement branch. 
Additionally, we compared models using multi-resolution versus single-resolution ``DEMUCS-32ms'' in Table~\ref{table-stationary}. 
Their performances were almost the same, suggesting that improving the performance with a single network output is difficult.

\begin{table*}[ht]
	\centering
	\small
	\caption{\small Results of different SE systems and the proposed method. ``Causal'' indicates that the model can be used for streaming operations.}
	\vspace{5pt}
	\begin{tabular}{lccccccc}
		\toprule
		System & segSNR ($\uparrow$) & CSIG ($\uparrow$) & CBAK ($\uparrow$) & COVL ($\uparrow$) & STOI ($\%$, $\uparrow$) & PESQ ($\uparrow$) & Causal \\
		\midrule
		Noisy & 1.68 & 3.35 & 2.44 & 2.63 & 91.5 & 1.97 & \ding{56} \\
		SEGAN \cite{Pascual2017} & 7.73 & 3.48 & 2.94 & 2.80 & -    & 2.16 & \ding{56}\\
		SEGAN-D \cite{9201348} & 8.72 & 3.46 & 3.11 & 3.50 & 93.3 & 2.39 & \ding{56} \\
		Wave U-Net \cite{DBLP:journals/corr/abs-1811-11307} & 9.97 & 3.52 & 3.24 & 2.96 & - & 2.40 & \ding{56} \\
		MMSE-GAN \cite{8462068} & - & 3.80 & 3.12 & 3.14 & 93.0 & 2.53 & \ding{56} \\
		MetricGAN \cite{pmlr-v97-fu19b} & - & 3.99 & 3.18 & 3.42 & - & 2.86 & \ding{56} \\
		S-DCCRN \cite{9747029} & - & 4.03 & 2.97 & 3.43  & 94.0 & 2.84 & \ding{52} \\
		DeepMMSE \cite{9066933} & - & 4.28 & 3.46 & 3.64 & 94.0 & 2.95 & \ding{56} \\
		PHASEN \cite{Yin_Luo_Xiong_Zeng_2020} & \textbf{10.18} & 4.21 & \textbf{3.55} & 3.62 & - & 2.99 & \ding{56} \\
		\midrule
		DEMUCS \cite{Dfossez2020} & 8.74 & 4.22 & 3.25 & 3.52 & 94.8 & 2.93 & \ding{52} \\
		\midrule
		DEMUCS-MRE (proposed) & 8.95 & 4.38 & \textbf{3.52} & 3.73 & \textbf{95.1} & 3.03 & \ding{52} \\
		DEMUCS-MRD (proposed) & \textbf{9.07} & 4.33 & 3.49 & 3.68 & 94.6 & 2.98 & \ding{52} \\
		DEMUCS-MRE-MRD (proposed) & 8.73 & \textbf{4.40} & \textbf{3.52} & \textbf{3.77} & \textbf{95.1} &  \textbf{3.07} & \ding{52} \\
		\bottomrule
	\end{tabular}
	\label{table1}
	\vspace{-10pt}
\end{table*}

\begin{table}[ht]
	\centering
	\small
	\caption{\small The enhancement performance of DEMUCS-MRE with non-stationary frequency information (32ms, 64ms, 128ms STFT points).}
	\vspace{5pt}
	\begin{tabular}{p{4.5cm}cc}
		\toprule
		System  & STOI ($\%$, $\uparrow$) & PESQ ($\uparrow$) \\
		\midrule
		{\footnotesize DEMUCS-MRE (8ms,16ms,32ms)}   & 95.1 & 3.03 \\
		{\footnotesize DEMUCS-MRE (32ms,64ms,128ms)} & 94.7 & 3.00 \\
		\bottomrule
	\end{tabular}
	\label{table3}
	\vspace{-10pt}
\end{table}

\begin{table}[ht]
	\centering
	\small
	\caption{\small Comparison different time-domain outputs in ``DEMUCS-MRD''.}
	\vspace{5pt}
	\begin{tabular}{p{1.8cm}ccc}
		\toprule		
		System & 32ms output & 64ms output  & 128ms output  \\
		\midrule
		STOI ($\%$, $\uparrow$) & 94.6 & 94.6 & 94.6 \\
		PESQ ($\uparrow$)            & 2.99 & 2.98 & 2.96 \\
		\bottomrule
	\end{tabular}
	\label{table5}
	\vspace{-10pt}
\end{table}

\subsection{Effect of Fusing Frequency Information in Encoder}
Table~\ref{table1} shows the results of different SE systems and the proposed method. 
With the proposed method of ``DEMUCS-MRE'', there is a 0.1 PESQ improvement when the redundant frequency-domain information is added to the time-domain encoder. 
CSIG, CBAK, and COVL improvements show that the proposed ``DEMUCS-MRE'' could maintain more speech signals, suppress more noise, and improve overall quality.

Furthermore, we also tried the same resolutions as the DEMUCS (32ms, 64ms, and 128ms STFT points) in DEMUCS-MRE. 
The results in Table~\ref{table3} show that the performance of this system is slightly degraded compared to stationary ``DEMUCS-MRE (8ms,16ms,32ms)''. 
This shows that processing stationary signals in the frequency domain is more effective. 
Nevertheless, fusing the non-stationary frequency information in the encoder can also significantly improve the model performance. 
It is often pointed out that the frequency-domain SE systems have more stable enhancement performance than time-domain SE systems \cite{Zhao2020} because the instability of the phase information makes the time-domain waveform less stable than the frequency-domain magnitude of the spectrogram. 
Incorporating frequency information into time-domain information can improve the stability of time-domain information.

\subsection{Effect of Multiple Decoders}
Table~\ref{table1} shows that the ``DEMUCS-MRD'' can provide 0.05 PESQ improvement. 
We averaged multiple outputs to get the final enhanced waveform. 
Table~\ref{table5} shows results of different time-domain outputs in ``DEMUCS-MRD''. 
The output of all different resolutions shows some improvement, 
which suggests that adopting multiple outputs can alleviate the problem of learning mismatch with a single output. 
The results of the different outputs were almost the same, especially for 32ms and 64ms. 
This may be because the model already has multi-resolution frequency domain information through the shared $1$-th to $4$-th decoder layers. 
Furthermore, ``DEMUCS-MRD'' achieved the best performance for segSNR, which means better performance at the segment level.

\subsection{Effect of Improving the Model with Both Encoder and Decoder}
The combination of the proposed MRE and MRD can provide further improvement, which is shown in the last row of Table~\ref{table1}. 
``DEMUCS-MRE-MRD'' has stronger speech signal retention and signal overall quality recovery ability. 
The PESQ improvement from the baseline ``DEMUCS'' is 0.14.

}

\section{Conclusions}
{
\small
In this paper, we proposed using multi-resolution encoders and decoders to solve the drawbacks of DEMUCS from both the encoder and decoder perspectives. 
We first added multi-resolution stationary frequency information to the time-domain enhancement layer by layer to solve the non-stationary STFT loss issue. 
The experimental results show that the stationary frequency information can significantly improve performance. 
Moreover, we adopted multiple time-domain outputs to alleviate the problem of learning mismatch with a single output.  
The results show that it can ensure the model has multi-resolution frequency information while improving the performance of all resolutions. 
Furthermore, the proposed MRE and MRD can be used jointly to achieve further improvement. 
In the future, we will fuse the multiple enhanced waveforms into one signal with a neural network instead of averaging. 
}

\section{Acknowledge}
This work was supported by JST, the establishment of university fellowships towards the creation of science technology innovation, Grant Number JPMJFS2123.

\footnotesize
\bibliographystyle{IEEEtran}
\bibliography{refs}

\end{document}